\documentclass[aps,nofootinbib,reprint,nobibnotes,showpacs,superscriptaddress]{revtex4-1} 

\usepackage[english]{babel}
   \addto\captionsenglish{
                         }

\usepackage{graphicx}
   \graphicspath{{graphics/}}

\usepackage[caption=false]{subfig}

\usepackage[shortlabels]{enumitem}

\usepackage{amsmath}
\usepackage{amssymb}
\usepackage[version=3]{mhchem}
\usepackage{mathrsfs}
\usepackage{bm}

\usepackage[hidelinks,linktocpage]{hyperref}
\hypersetup{
            colorlinks,
            linkcolor={red!70!black},
            citecolor={green!70!black},
            urlcolor={blue!70!black}
           }

\usepackage[usenames,dvipsnames]{xcolor}
\usepackage{comment} 
\usepackage{slashed} 
\usepackage{cancel}
\usepackage[normalem]{ulem}

\newcommand{\bb}{\ensuremath{0\nu\beta\beta}} 

\newcommand{\taubb}{T_{1/2}} 

\begin{document}

   \title{Complementarity between neutrinoless double beta decay and collider searches 
   \\
   for heavy neutrinos in composite-fermion models}

   \author{S.\ Biondini}
     \email{simone.biondini@unibas.ch}
      \affiliation{Department of Physics, University of Basel, Klingelbergstr. 82, CH-4056 Basel, Switzerland}
      \author{S.\ Dell'Oro}
      \affiliation{University of Milano-Bicocca, I-20126 Milano, Italy}
      \affiliation{INFN Sezione di Milano-Bicocca, I-20126 Milano, Italy}
   \author{R.\ Leonardi}
      \affiliation{INFN, Sezione di Perugia, I-06123 Perugia, Italy}
   \author{S.\ Marcocci}
      \affiliation{Fermi National Accelerator Laboratory, Batavia, Illinois 60510, USA}
   \author{O.\ Panella}
      \affiliation{INFN, Sezione di Perugia, I-06123 Perugia, Italy}
   \author{M.\ Presilla}
      \affiliation{INFN, Sezione di Perugia, I-06123 Perugia, Italy}
   \author{F.\ Vissani}
      \affiliation{INFN, Laboratori Nazionali del Gran Sasso, I-67100 Assergi, L’Aquila, Italy}

   \date{\today}
   \begin{abstract}
      Composite-fermion models predict excited quarks and leptons with mass scales which can potentially be observed at high-energy colliders like the LHC;
      the most recent exclusion limits from the CMS and ATLAS Collaborations corner excited-fermion masses and the compositeness scale to the multi-TeV range.
      At the same time, hypothetical composite Majorana neutrinos would lead to observable effects in neutrinoless double beta decay (\bb) experiments.
      In this work, we show that the current composite-neutrino exclusion limit $M_N>4.6$ TeV, as extracted from direct searches at the LHC, 
      can indeed be further improved to $M_N>8.8$ TeV by including the bound on the nuclear transition \ce{^{136}Xe \to ^{136}Ba + 2e^-}.
      Looking ahead, the forthcoming HL-LHC will allow probing a larger portion of the parameter-space, nevertheless, it will still benefit from the complementary limit provided by \bb\ future detectors
      to explore composite-neutrino masses up to $12.6$ TeV.
   \end{abstract}

   \maketitle


    Composite-fermions scenarios offer a possible solution to the hierarchy pattern of fermion masses~\cite{Pati:1975md,Pati:1983dh,Harari:1982xy,Greenberg:1974qb,Dirac:1963aa,Terazawa:1976xx}.
    The main phenomenological consequences of this class of models are the existence of heavy excitations of the Standard Model (SM) fermions, i.\,e.\ of excited quarks and leptons -- a hypothesis that is indeed tested in high-energy experiments -- and of gauge and contact interactions between SM fermions and excited fermions~\cite{ Eichten:1979ah,Eichten:1983hw,Cabibbo:1983bk,Baur:1989kv,Baur:1987ga,Olive:2014aa}.
    The excited states are expected to have masses ranging from the electroweak~\cite{Eichten:1979ah,Cabibbo:1983bk,Pancheri:1984sm} up to the compositeness scale and can be embedded in weak-isospin multiplets, thus coupling to the ordinary fermions via gauge interactions with magnetic-type transition~\cite{Cabibbo:1983bk,Pancheri:1984sm}.
    
    In this work, we probe a class of composite-fermion models by exploiting the complementarity between the direct searches at high-energy colliders and phenomenological manifestations at a much lower energy scale, in particular in neutrinoless double beta decay reactions.
    
    Neutrinoless double beta decay (\bb) is a rare nuclear process forbidden by SM that violates the lepton number by two units; its observation would demonstrate that the lepton number is not a symmetry of nature.
    The theoretical framework preferred by the community sees the \bb\ transition mediated by the exchange of ordinary, light neutrinos. As a matter of fact, we have proven the existence of a non-zero neutrino mass, while at the same time the structure of the SM would be minimally extended by including a Majorana mass term for the neutrino~\cite{Dell'Oro:2016dbc}.
    Nevertheless, alternative mechanisms can be invoked to explain the \bb\ process, such as the exchange of composite heavy Majorana neutrinos.

    The investigation of composite-fermion scenarios has recently been the object of phenomenological studies and experimental analyses at the Large Hadron Collider (LHC) searching for excited
   quarks~\cite{CMS:2017kmn,ATLAS:2015esi,CMS:2015xau},
   charged leptons~\cite{atlas-limit,atlas-limit-new,cms-limit-7TeV,cms-limit-8TeV,cms-limit-13TeV,CMS-PAS-EXO-18-004,Sirunyan:2018zzr,CMS-PAS-EXO-18-013} and, indeed,
   Majorana neutrinos~\cite{Leonardi:2015qna,Sirunyan:2017xnz,CMS-PAS-FTR-18-006,CidVidal:2018eel}.
   At the same time, the cosmological implication of the neutral composite leptons has been explored in the context of baryogenesis via leptogenesis~\cite{Zhuridov:2016xls,Biondini:2017fut}. 
   These studies were based on the following assumptions~\cite{Nakamura:2010zzi}:
   
   \textit{(a)} the charged current that involves SM gauge bosons and the excited Majorana neutrino $\nu^* \equiv N$ is of magnetic type, i.\,e.\ it is described by a dimension-5 operator:
         \begin{equation}
         \label{lagGI}
            \mathcal{L}_{\text{GI}}=\frac{gf}{\sqrt{2}\Lambda}\, \bar{N} \, \sigma_{\mu \nu} \, \ell_L \,\,  \partial^{\nu} \, W^{\mu} \,  + h.\,c \, ,
         \end{equation}
         where $g$ is the SU(2) SM gauge coupling, $\Lambda$ is the compositeness scale, $f$ is a free parameter of the model and $\sigma^{\mu \nu}=i\left[ \gamma^{\mu}, \gamma^{\nu}\right]/2$;
         
    \textit{(b)} contact interactions between ordinary fermions may arise by the exchange of more fundamental constituents, if these are commons to fermions,
         and/or by the exchange of the binding quanta of the new unknown interaction~\cite{Peskin:1985symp,Olive:2014aa}.
         The dominant effect is expected to be given by a dimension-6 operator, namely four-fermion interactions scaling with the inverse square of the compositeness scale: 
         \begin{equation}
         \label{Lcontact}
               \mathcal{L}_{\text{CI}} =\frac{g_\ast^2}{2 \Lambda^2}j^\mu j_\mu \, .
         \end{equation}
         The effective strong coupling $g_*$ is analogous to the $\rho$-meson effective coupling $g_\rho^2/(4\pi) \approx 2.1$
         arising from the new ``meta-color'' force exchanged between preon sub-constituents; it is normalised, according to standard implementations,
         by setting $g_*^2 = 4\pi$~\cite{Olive:2014aa,Biondini:2014dfa}.
         The current $j_\mu$ is actually the sum of various vector/axial-vector currents:
         \begin{multline}
         \label{Jcontact}
           j_\mu = \eta_L\bar{f_L}\gamma_\mu f_L+\eta\prime_L\bar{f^\ast_L}\gamma_\mu f^\ast_L \\
                   + \eta\prime\prime_L\bar{f^\ast_L}\gamma_\mu f_L + h.\,c. + (L\rightarrow R).
         \end{multline}

   \begin{figure*}[t!]
      \subfloat[Gauge interactions.\label{fig:diag_GI}]{\includegraphics[width=.24\linewidth]{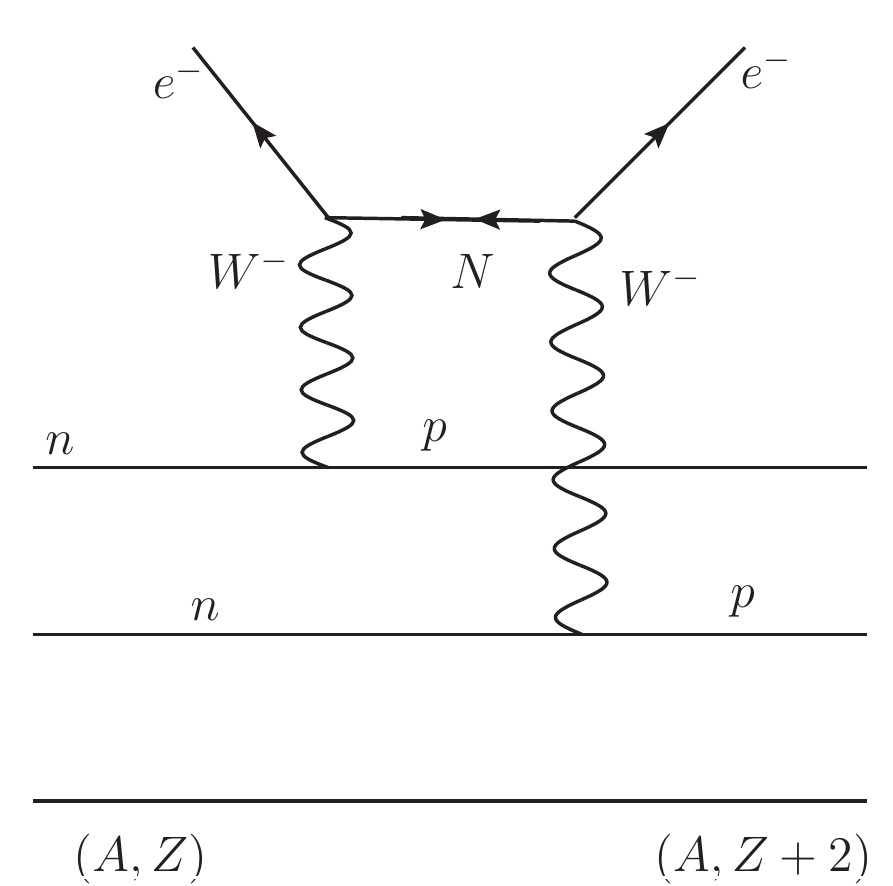}}
      \subfloat[Contact interactions.\label{fig:diag_CI}]{\includegraphics[width=.24\linewidth]{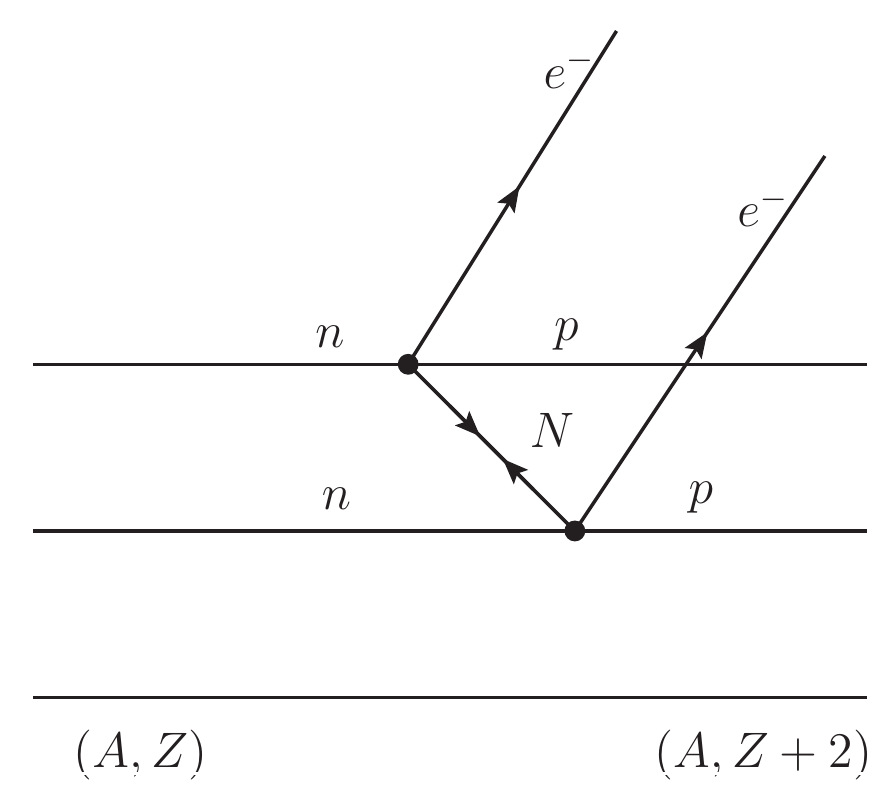}}
      \subfloat[Mixed term I.\label{fig:diag_MI1}]{\includegraphics[width=.24\linewidth]{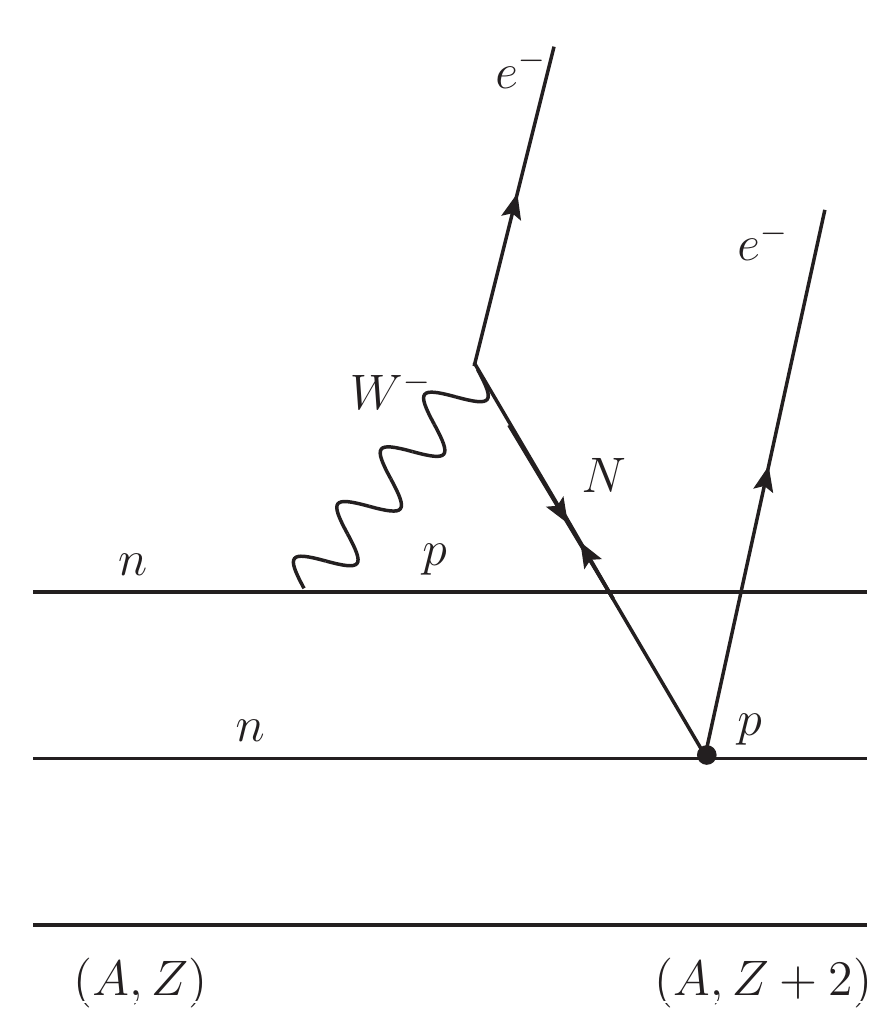}}
      \subfloat[Mixed term II.\label{fig:diag_MI2}]{\includegraphics[width=.24\linewidth]{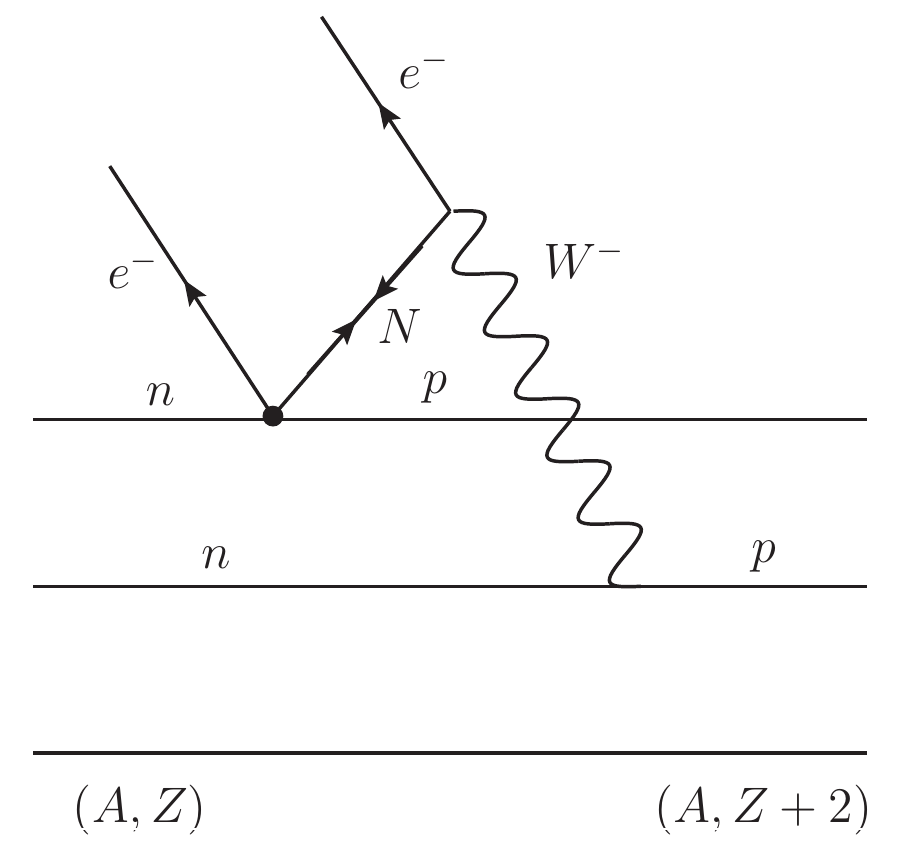}}
      \caption{Feynman diagrams for \bb, i.\,e.\ the transition of two neutrons $n$ into two protons $p$, mediated by the exchange of heavy composite neutrinos $N$ within a composite-fermion model.
      The two vertices involve (from left to right), magnetic-type interactions from Eq.~\eqref{lagGI},
       contact interaction vertices from Eq.~\eqref{Lcontact2} and one gauge and one contact interaction vertex (two permutations).
      In the text, we refer to them as pure gauge (a), pure contact (b) and mixed contributions (c) and (d).}
   \label{fig:diagrammi}
   \end{figure*}

   In this work, right-handed currents are neglected for simplicity, as commonly adopted by the collider community.
   Flavour conserving but non-diagonal terms, in particular those with currents like the third term in Eq.~\eqref{Jcontact}, can couple excited states with ordinary fermions,
   so that Eq.~\eqref{Lcontact} contains a term of  the form:
   \begin{equation}
   \label{Lcontact2}
      \mathcal{L}_{\text{CI}}=  \frac{\eta_L \,g_\ast^2}{\Lambda^2} \, \left(\sum_{q,q'}\,\bar{q}_L\gamma^\mu  q_L'\right) \, \bar{N}_L\gamma_\mu \ell_L  + h.\,c.\, ,
   \end{equation} 
   when selecting charged SM leptons accompanying the heavy exited neutrino. We shall not distinguish between the model parameters $\eta$'s in Eq.~\eqref{Jcontact},
   and simply indicate them all with a generic $\eta_L$. These interactions can account for the production of excited neutrinos at hadron colliders via the $2\to2$ process $q\bar{q}' \to N\ell$, as recently shown in phenomenological studies~\cite{Leonardi:2014aa,Leonardi:2015qna}.

   As we will show, the contact interactions induce \bb, and actually provide the dominant contribution when compared to the gauge interactions for this model. 
   The relevant diagrams for \bb, that involve a composite Majorana neutrino with gauge and contact interactions, are illustrated in Fig.~\ref{fig:diagrammi}.
   In particular, the contribution by Eq.~\eqref{lagGI} (Fig.~\ref{fig:diag_GI}) had already been calculated in Refs.~\cite{Panella:1995aa,Panella:1997aa,Panella:1998aa}.
   Here, we calculate the additional contribution due to contact interactions, as expressed in Eq.~\eqref{Lcontact2} (Fig.~\ref{fig:diag_CI}) and estimate the effect from the mixed terms
   (Figs.~\ref{fig:diag_MI1} and \ref{fig:diag_MI2}).

   We can rewrite the Lagrangian of Eq.~\eqref{Lcontact2}, which describes the four-fermion contact interactions, as follows:
   \begin{equation}
   \label{Lcontact3}
      \mathcal{L}_{\textrm{CI}}=\frac{g_*^2}{2 \Lambda^2}J_\mu J_h^\mu \, ,
   \end{equation}
   where $J_\mu=\bar{\Psi}_e \gamma_\mu \eta_L P_L \Psi_N$, with $P_{L} \equiv (1-\gamma^5)/2$, while $J^h_\mu=\sum_{q,q'}\,\bar{q} \gamma_\mu (1-\gamma^5)  q'$ is the hadronic weak charged current
   induced by the quark-level current. For the latter current, we factor out $1/2$ from the chiral projector $P_L$ in Eq.~\eqref{Lcontact3},
   in order to conform with the common expressions of the hadronic current and nuclear matrix elements~\cite{Doi:1983aa,Haxton:1984aa,Vergados:2012aa,Engel:2016xgb}.
   The corresponding $S$-matrix element is
   \begin{eqnarray}
   \label{eq:Smatrix}
      S_{\textrm{CI}} & = & \left(\frac{g_*^2}{\Lambda^2}\right)^2 \frac{1}{8} \int \frac{d^4q}{(2\pi)^4}d^4xd^4y \; e^{-iq\cdot(x-y)} \nonumber \\ 
      &\times& \frac{\eta_L^2}{\sqrt{2}}(1-P_{12})\bar{\Psi}(p_2)  \gamma_\mu \, P_L \, \frac{q\mkern-9mu \slash+M_N}{q^2-M_N^2}  P_L \, \gamma_\nu \Psi^c(p_1)  \nonumber \\
      & \times & e^{i(p_1\cdot x+p_2\cdot y)} \langle F | T[J^\mu_h(x)J^\nu_h(y)]|I\rangle \, ,
   \end{eqnarray}
   where $(1-P_{12})/\sqrt{2}$ is the antisymmetric operator due to the production of two identical fermions, (two electrons in our case) and $\Psi^c=C \bar{\Psi}^T$,
   where $C$ is the charge conjugation matrix.

   We make the \emph{ansatz} that the hadronic current is given by the corresponding sum of the nucleonic charged current~\cite{Lipkin:1955aa,Tamura:1956aa,ring:1980:aa}, namely $J^h_\mu(x)=\sum_i J^{(i)}_\mu(x) $,
   where the sum runs over the nucleons of the isotope which undergoes \bb. Therefore, we can rewrite
   \begin{eqnarray}
      \langle F|T[J_h^\mu(x)J_h^\nu(y)]|I\rangle &=&\exp{[i(p_F-p_I)\cdot y]} \nonumber \\
      && \times \langle F|T[J_h^\mu(x-y)J_h^\nu(0)]|I \rangle\, ,\phantom{xxxx}
   \end{eqnarray}
   where $p_{F(I)}$ refers to the outgoing (incoming) hadron momentum. 
   We change the integration variables as $x=z+u/2$ and $y=z-u/2$ in Eq.~\eqref{eq:Smatrix}, so to obtain 
   \begin{eqnarray}
      S_{\textrm{CI}} & = & \left(\frac{g_*^2}{\Lambda^2}\right)^2\frac{1-P_{12}}{8\sqrt{2}}\int\frac{d^4q}{(2\pi)^4}d^4zd^4u \; e^{-iq\cdot u}\nonumber \\
      &\times & \, e^{iz\cdot(p_1+p_2+p_F-p_I)} \;  \eta_L^2  M_N \bar{\Psi}(p_2) \gamma_\mu\gamma_\nu \, P_R \Psi^c(p_1)  \nonumber \\
      & \times&\,  e^{i(u/2)\cdot(p_1-p_2-p_F+p_I)} \,  \frac{\langle F|T[J_h^\mu(u)J_h^\nu(0)]|I\rangle}{(q^2-M_N^2)}.
   \end{eqnarray}
   The integration over $z$ guarantees the energy-momentum conservation, and we recast the matrix element in the form $S_{\textrm{CI}}=(2\pi)^4\delta^4(p_I-p_F-p_1-p_2)T_{\textrm{CI}}$, where:
   \begin{eqnarray}
   \label{eq:Tfi}
      T_{\textrm{CI}}  &=& \eta_L^2 \left(\frac{g_*^2}{\Lambda^2}\right)^2\frac{(1-P_{12})}{8 \sqrt{2}}\int\frac{d^4q}{(2\pi)^4} \, d^4u \; e^{-i(q-p_1)\cdot u}\nonumber \\ 
      &\times&  M_N\bar{\Psi}(p_2)\gamma_\mu\gamma_\nu P_R \, \Psi^c(p_1) \, \frac{\langle F|T[J_h^\mu(u)J_h^\nu(0)]|I\rangle}{(q^2-M_N^2)}\,.\phantom{xxx}
   \end{eqnarray}
   The leptonic current can be simplified with standard Dirac algebra, and by defining
   \begin{eqnarray}
   \label{Wmunuq}
      W^{\mu\nu}(q) \equiv \int d^4x \, e^{-iq\cdot x}\langle F|T[J_h^\mu(x)J_h^\nu(0)]|I\rangle \, ,
   \end{eqnarray}
   we can write Eq.~\eqref{eq:Tfi} as follows:
   \begin{eqnarray}\label{eq:b}
      T_{\textrm{CI}}=&&\left(\frac{g_*^2}{\Lambda^2}\right)^2\frac{\eta_L^2 M_N}{4\sqrt{2}}\bar{\Psi}(p_2)\, P_R \,\Psi^c(p_1)\nonumber \\
      && \times \int\frac{d^3 \bm{q}}{(2\pi)^3}\int\frac{dq_0}{2\pi}\frac{W^\mu_{\phantom{\mu}\mu}(q-p_1)}{(q_0^2-\omega_N^2+i\epsilon)} \, ,
      \label{Tfi_W}
   \end{eqnarray}
   where $\omega_N=\sqrt{\bm{q}^2+M_N^2}$. 
   Following Ref.~\cite{Panella:1997aa}, we expand $W_{\mu\nu}$ in Eq.~\eqref{Wmunuq} by using a complete set of intermediate states 
   and notice that the energy of a state $|X\rangle$ can be written as $E_X=E_{c.m.}(\bm{P})+\epsilon_X$, where $E_{c.m.}(\bm{P})$ is the energy of the center of mass motion and $\epsilon_n$
   is the excitation energy.
   As commonly performed in \bb\ calculations, we use the closure approximation, i.\,e.\ we replace the energy of the intermediate state $E_X$ with an average value $\langle E_X\rangle= E_{\text{c.m.}}\left(\langle \bm{P}_{\text{X}}\rangle\right)+ \bar{\epsilon}_X$,  where $\bar{\epsilon}_X$ is the average excitation energy of the intermediate states~\cite{Haxton:1984aa,Burgess:1994aa,Pantis:1990ai,Pantis:1996py} and is typically of the order of 10 MeV. 
   The virtual neutrino momentum $|\bm{q}|$ (equal to the momentum transfer in the process) is of the order of $|\bm{q}|\approx 1/r_{NN}$ where $r_{NN}\approx 2$\,fm
   is the average inter-nucleon distance in the nuclei so that $ |\bm{q}| \approx 100$ MeV $\ll M_{N}$. 
   This means that the energy of the center-of-mass motion of the nuclei is negligible relative to the typical excitation energies (10 MeV) and also to the initial
   and final nuclei energies ($E_I, E_F$), so that $E_I\approx M_I$ and $E_F\approx M_F$.
   By integrating over the center-of-mass momentum of the intermediate state, and by introducing the so called closure energy~\cite{Haxton:1984aa,Tomoda:1991aa,Barea:2015aa}
   \begin{equation}
      \Delta\approx E_{\text{c.m.}}\left(\langle \bm{P}_{\text{X}}\rangle\right)+ \bar{\epsilon}_X-\frac{1}{2}(M_I+M_F) \approx 10 \text{\ MeV} \, ,
   \end{equation}
   we obtain the tensor $W^{\mu\nu} (q-p_1)$ as:
   \begin{eqnarray}
   \label{Wmunuclosure1}
       W^{\mu \nu}(q^0, \bm{q}) &=& i \frac{2  \Delta}{q_0^2-\Delta^2 + i \epsilon} \\
                                &\times& \langle \langle F | \sum_{n n'}\exp \left( i \bm{q} \cdot \bm{r}_{nn'} \right) \, J_n^\mu(-\bm{q})J^\nu_n(\bm{q})| I \rangle \rangle \nonumber\, ,
   \end{eqnarray}
   where $\bm{r}_{nn'}= \bm{r}_n- \bm{r}_n'$ is the nucleons' relative position vector, $J^\mu(\bm{q})$ is the nucleon current in momentum space and $\langle\langle F| \cdots | I\rangle \rangle$
   denotes the matrix element over the $A-1$ relative coordinates once the center of mass motion has been integrated out.
   Notice that the dependence from $p_1$ in Eq.~\eqref{Wmunuclosure1} is marginal, and we drop it in the following,  because: 
   (i) the momentum of the final electron $\bm{p_1}\approx 1$ MeV can always be neglected relative to the virtual neutrino momentum $\bm{q} \approx 100$ MeV;
   (ii)    the energy of the two final electrons $E_1+E_2$ in \bb\ is fixed to approximately 2 MeV, and the energy  of the final electrons $E_{1,2}\lessapprox 2$ MeV  is fairly smaller than the average excitation energy $\Delta$.

   Out of the various available formulations for the non-relativistic nucleon currents and corresponding normalizations, we consider the ones given in Refs. \cite{Simkovic:1999re,Barea:2013bz}:
   \begin{eqnarray}
      &&J^{(n)}_0 (\bm{q}) = g_V(\bm{q}^2) \tau^+_{n}\, , \label{J0_NR} \\
      &&J^{(n)}_i (\bm{q}) = \left[ g_A(\bm{q}^2) (\bm{\sigma}_{n})_i - g_P(\bm{q}^2) \frac{\bm{\sigma}_n \cdot \bm{q}}{2 m_p} q_i \right. \label{Ji_NR} \nonumber \\
      &&\left. \hspace{2 cm} + i g_M(\bm{q}^2) \frac{(\bm{\sigma}_{n} \times \bm{q})_i}{2 m_p} \right] \tau^+_{n} \, ,
   \end{eqnarray}
   where $\bm{\sigma}_k $ is the spin matrix of the ${k}$-th nucleon, labelled with $n$, and $\tau^+_{n}$ is the ladder operator of the nuclear isospin. The values of the form factors
   $g_V(\bm{q}^2),g_A(\bm{q}^2), g_P(\bm{q}^2)$ and $g_M(\bm{q}^2)$, and relevant parameters in Eqs.~\eqref{J0_NR} and \eqref{Ji_NR} are fixed as in Ref.~\cite{Barea:2013bz},
   and we specify here the two form factors that act as building blocks for the remaining ones
   \begin{equation}
       g_A(\bm{q}^2)=\frac{g_A}{(1+\bm{q}^2/ M_A^2)^2} \, , \quad  g_V(\bm{q}^2)=\frac{g_V}{(1+\bm{q}^2/ M_V^2)^2}  \, ,
   \end{equation}
   where  $g_V = 1$ (under the hypothesis of conserved vector current), $M_V^2 =0.71$ $(\textrm{GeV}/c^2)^2$~\cite{Dumbrajs:1983jd},
   $g_A \simeq 1.269$~\cite{ParticleDataGroup:2006fqo} and $M_A=1.09$ ($\textrm{GeV}/c^2)^2$~\cite{Schindler:2006jq}. 
   
   Finally, the quantity $W^\mu_{\phantom{\mu}\mu} (q-p_1)$ appearing in Eq.~\eqref{Tfi_W} (within the closure approximation) is given by
   \begin{eqnarray}
      W^{\mu}_{\phantom{\mu} \mu}(q^0,\bm{q}) &=& -i \frac{2  \Delta}{q_0^2-\Delta^2 + i \epsilon} \\
                                              &\times& \langle \langle F | \sum_{n n'}\exp \left( i \bm{q} \cdot \bm{r}_{nn'} \right) \, \Omega_{n n'}(\bm{q})| I \rangle \rangle , \nonumber
   \end{eqnarray}
   with the two body effective transition operator in momentum space of the form~\cite{Simkovic:1999re}:
   \begin{eqnarray}
      \Omega_{n n'}(\bm{q}) &=& \tau^+_n  \tau^+_{n'}  \left[ -h_\textrm{F}(q) + h_{\textrm{GT}}(q) \bm{\sigma}_n \cdot \bm{\sigma}_{n'} \phantom{xxxxxxx}\right. \nonumber \\
                            &&\phantom{xxxxxxxxxxxxxxxxx}\left. - h_{\textrm{T}}(q) \, S_{n n'} \right] \, ,
   \end{eqnarray}
   with $S_{nn'}=3\left[(\hat{\bm{q}} \cdot \bm{\sigma}_n )(\hat{\bm{q}} \cdot \bm{\sigma}_{n'} )\right] - \bm{\sigma}_n \cdot \bm{\sigma}_{n'}$, and the functions
   $h_\textrm{F}(q)$, $ h_{\textrm{GT}}(q)$ and $h_{\textrm{T}}(q)$ can be found in e.g.~Refs.\cite{Simkovic:1999re,Barea:2013bz}. 
   Performing the integration upon the temporal component of the momentum transfer ($q^0$) in Eq.~\eqref{Tfi_W} we define, and calculate, the integral:
   \begin{eqnarray}
   \label{result_I0}
      I(\bm{q}^2) &\equiv& \int\frac{dq_0}{2\pi i}\left[\frac{\Delta}{q_0^2-\Delta^2+i\epsilon} \right]\frac{1}{(q_0^2-\omega_N^2+i\epsilon)} \nonumber \\
                  &=& - \frac{1}{2\omega_N(\omega_N+\Delta)} \, .
   \end{eqnarray}
   According to the assumed heavy-neutrino mass and kinematic of interest, we take $\omega_N \approx M_N$ and we can safely drop powers of $\Delta/M_N$ in Eq.~\eqref{result_I0}.
   Then, we obtain the following expression for $T_{\textrm{CI}}$
   \begin{eqnarray}
   &&   T_{\textrm{CI}}  =  \left(\frac{g_*^2}{\Lambda^2}\right)^2\frac{\eta_L^2}{4\sqrt{2}}\bar{\Psi}(p_2) P_R \Psi^c(p_1)  
   \nonumber \\
    && \, \int\frac{d^3\bm{q}}{(2\pi)^3}   \langle \langle F | \sum_{n n'}\exp \left( i \bm{q} \cdot \bm{r}_{nn'} \right) \, 
    \Omega_{n n'}(\bm{q}) | I \rangle \rangle  \,.
   \end{eqnarray}
   Next we can express the result in terms of standard nuclear matrix elements (NMEs), for a heavy Majorana neutrino exchange,  as follows \cite{Simkovic:1999re,Barea:2013bz}
   \begin{equation}
   \label{eq:Tfi_6}
      T_{\textrm{CI}} = \left(\frac{g_*^2}{\Lambda^2}\right)^2\!\!\!\frac{\eta_L^2}{4\sqrt{2}}\bar{\Psi}(p_2) P_R \Psi^c(p_1)
         \frac{m_p m_e}{ M_N} \frac{g_A^2}{4 \pi R_0} \mathcal{M}^{0N} , \phantom{xxx}
   \end{equation}
   where $R_0=r_0A^{1/3}$ is the mean nuclear radius, with $r_0=1.1$ fm, $m_e$ and $m_p$ are the electron an proton mass respectively, and the NME reads
   \begin{equation}
      \mathcal{M}^{0N}  =   \mathcal{M}^{0 N}_{\textrm{GT}} - \left( \frac{g_V}{g_A} \right)^2 \mathcal{M}^{0N}_{\textrm{F}} + \mathcal{M}^{0N}_{\textrm{T}} \, .
   \end{equation}
   Eq.~\eqref{eq:Tfi_6} enters the definition of the \bb~half-life, which is the actual observable from the experimental searches, as:
   \begin{multline}
   \label{eq:inv_taubb}
      [\taubb]^{-1}_{\textrm{CI}} = \frac{1}{\log 2}\int \frac{d^3\bm{p}_1}{(2\pi)^32E_1}\frac{d^3\bm{p}_2}{(2\pi)^32E_2} \\
      \overline{|T_{\textrm{CI}}|^2}\,2\pi\delta(E_I-E_F-E_1-E_2) \, ,
   \end{multline}
   where the amplitude squared, and summed over the electron spin polarization, is
   \begin{eqnarray}
      \overline{|T_{\textrm{CI}}|^2}&=&\left(\frac{g_*^2}{\Lambda^2}\right)^4\frac{\eta_L^4 \, g_A^4 (m_e \, m_p)^2}{512 \  \pi^2R_0^2M_N^2}|\mathcal{M}^{0N} |^2 
      \nonumber
      \\
      &&\hspace{1 cm}\sum_{\substack{\text{spin} \, e_1 e_2}}|\bar{\Psi}(p_2) \, P_R \, \Psi^c(p_1)|^2 .
   \end{eqnarray}
   In particular, the electron wave function $\Psi$ can be factorized in the following way, $\Psi(p)=\sqrt{F_0(Z+2,E)}\,u(p)$,
   where $F_0$ is the Fermi function describing the distortion of the electron wave in the Coulomb field of the nucleus~\cite{Doi:1983aa,Kotila:2012zza}. The spinor algebra simplifies then to:
   \begin{eqnarray}
      \sum_{\substack{\text{spin} \, e_1 e_2}}&&|\bar{\Psi}(p_2) P_R \Psi^c(p_1)|^2 =
      \nonumber \\
      &&= F_0(Z+2,E_1)F_0(Z+2,E_2) 2p_1\cdot p_2 \, . 
   \end{eqnarray}
 
   \begin{figure*}[t!]
      \includegraphics[width=1.\columnwidth]{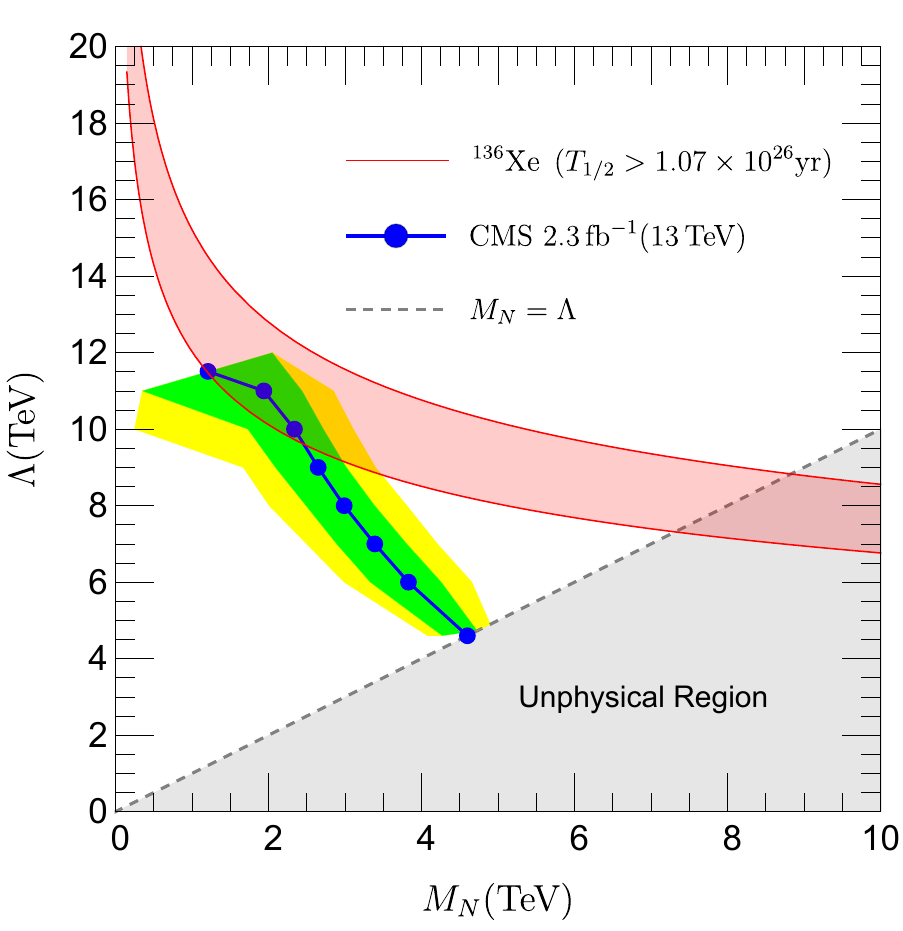} \quad
      \includegraphics[width=0.99\columnwidth]{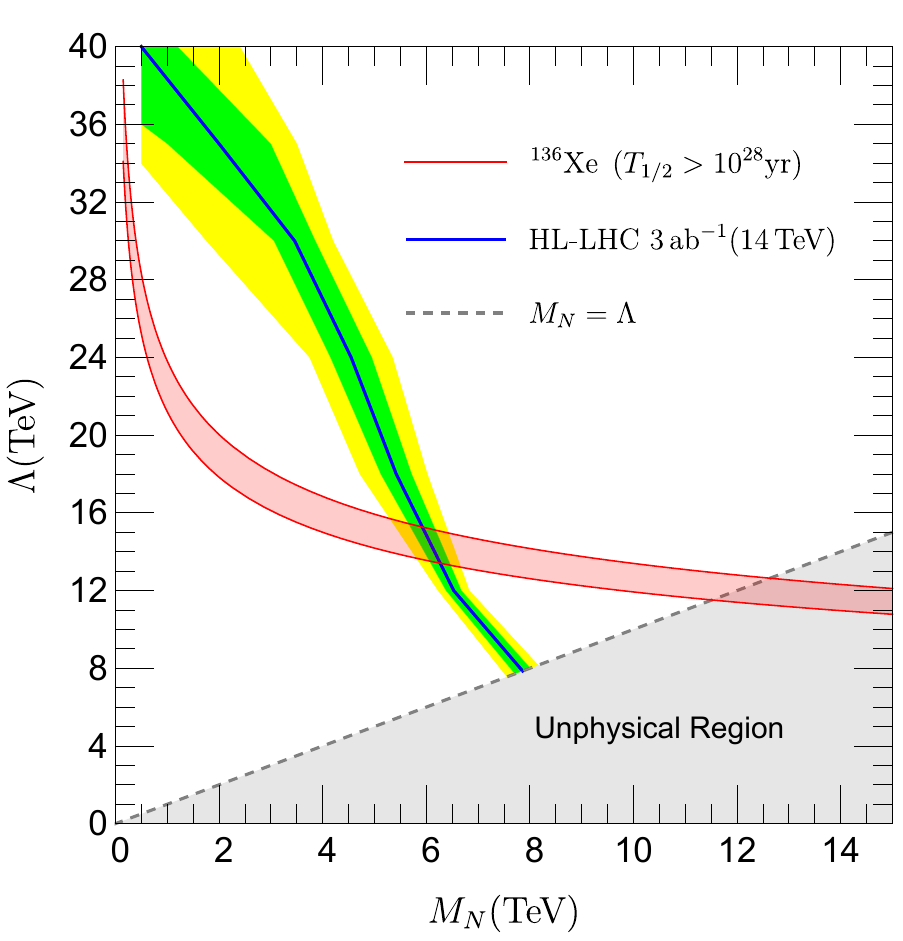}
      \caption{Lower bounds on the compositeness scale $\Lambda$ as function of the heavy Majorana neutrino mass $M_N$.
         (Left panel) The red semitransparent band is the bound from the $0\nu\beta\beta$ as given by the CI induced half-life Eq.~\eqref{NEWbound}, with the experimental value
         $T_{1/2}>1.07 \times 10^{26}$ yr~\cite{KamLAND-Zen:2016pfg}. The red lines correspond to the minimum and maximum values for the NMEs~\cite{Barea:2015kwa,Hyvarinen:2015bda.}
         The solid-dotted blue line is the bound from the analysis of 2.3 fb$^{-1}$ of data collected at the LHC during Run-2 at $\sqrt{s}=13$ TeV,
         by the CMS Collaboration~\protect\cite{Sirunyan:2017xnz}.
         The dashed-gray line, as corresponding to $\Lambda=M_N$, delimits the unphysical region for the model. (Right panel) Projection of the \bb\ bound with a half-life $T_{1/2} > 10^{28}$, the
         solid blue line stands for the CMS Collaboration projection study of the heavy composite neutrino at the HL-LHC \cite{CMS-PAS-FTR-18-006,CidVidal:2018eel}. The green and yellow bands correspond to the expected one and two standard 
         deviation(s), respectively.}
   \label{FigBound}
   \end{figure*}

   We adopt the standard notation~\cite{Kotila:2012zza} to express the phase-space integration and define 
   \begin{eqnarray}
      A_{0\nu}&=&\frac{(G_F\cos{\theta_c})^4m_e^9}{64\pi^5} \, , \\
      G_{01}&=&\frac{A_{0\nu}}{\ln2(m_eR_0)^2}\int\frac{2p_1E_1p_2E_2}{m_e^5}\delta(E_I-E_F-E_1-E_2) \nonumber \\
      &\times&F_0(Z+2,E_1)F_0(Z+2,E_2)dE_1dE_2  \, ,
   \end{eqnarray}
   where $G_{01}$ is the integral Phase Space Factor.
   Therefore, we can rewrite Eq.~\eqref{eq:inv_taubb} as:
   \begin{equation}
   \label{T_CI_final}
      \left[\taubb^{\textrm{CI}}\right]^{-1} = \left(\frac{g_*^2}{\Lambda^2}\right)^4\frac{\eta_L^4 \, g_{A}^{4}  \, m_p^2 }{64
   \, M_N^2}|\mathcal{M}^{0N}|^2\frac{G_{01}}{(G_F\cos\theta_c)^4 }  .\phantom{xxxxx}
   \end{equation}
   
   Eq.~\eqref{T_CI_final} represents the main analytical result of the paper. It complements the former finding from Ref.~\cite{Panella:1997aa},
   where the half-life from the sole gauge interactions had been derived.
   Notice that the new expression conforms with the general structure for the \bb~half-life as induced by a heavy-neutrino exchange~\cite{Doi:1983aa}. 
   
   Finally, a few considerations on the mixed diagrams are in order.
    The amplitude of the contributions in Figs.~\ref{fig:diag_MI1} and \ref{fig:diag_MI2} can be recast in terms of the CI amplitude $T_{\textrm{CI}}$ and of the ratios of the relevant scales,
    namely those appearing in the model and those typical of the nuclear dynamics.
    The overall term then takes the form
    \begin{multline}
    \label{MIX_approx}
      T_{\textrm{Mix}} \simeq i \frac{\cos (\theta_c)}{\sqrt{2}} \, \frac{g }{g_*^2} \, \frac{\Lambda}{M_N} \, \frac{\Delta}{M_W}  T_{\textrm{CI}} \\
      \approx i \, \, 6 \times 10^{-6} \; \frac{\Lambda}{M_N}  T_{\textrm{CI}}\, ,
   \end{multline}
   where we neglected the factor $|\bm{q}^2|/M_W^2$ appearing in the non-relativistic nuclear currents, which are further suppressed with respect to $\Delta/M_W$.

   The contribution from Eq.~\eqref{MIX_approx} could be potentially comparable in size with that from pure contact interactions, if one explored compositeness scales much larger than the
   composite-neutrino mass $M_N$.
   However, for values of $\Lambda$ not larger than 200--300 $M_N$, we can safely neglect the mixed diagrams at the amplitude level (refer to the plots in Fig.~\ref{FigBound});
   this holds even more when considering \bb\ half-lives, because the relevant quantity in this case becomes the square of the amplitude.
   At the same time, since Eq.~\eqref{MIX_approx} contains an unbalanced imaginary unit, whose presence is due to the odd number of $\sigma^{\mu \nu}$ entering the mixed diagrams,
   there is no contribution from the interference terms with the pure gauge and pure contact diagrams in Fig.~\ref{fig:diag_GI} and Fig.~\ref{fig:diag_CI} respectively.
   
   It is worth mentioning that the pure gauge contribution \cite{Panella:1997aa} is much smaller than the pure contact one.
   The suppression is induced by the ratio of the coupling combination $(g/g_*)^4$, as well as $m_e m_p/M_W^2$.
   Similarly to what happens for the mixed diagrams, there is an enhancing factor, here $(\Lambda/M_W)^2$, that may compensate for the suppression factors only for $\Lambda \approx 10^6$ GeV,
   which is far beyond the parameter range of interest in our work. For the same reason, the interference between the pure gauge and contact diagrams is also negligible; we actually performed a numeric verification of the above-mentioned consideration. Therefore, we will focus the following discussion on the pure-contact contribution.

   Up to date, the most stringent experimental bounds on \bb\ come from the searches 
   \begin{equation}
      \ce{^{76}Ge \to ^{76}Se + 2e^-} \mbox{ and } \ce{^{136}Xe \to ^{136}Ba + 2e^-} \, ,
   \end{equation}
   where the limit on the decay half-life are
   \begin{equation}
     \taubb ~ (\mbox{90\%\,C.\,L.}) > \left\{
     \begin{aligned}
        &1.8 \times 10^{26} \mbox{yr~~(\ce{^{76}Ge},~\cite{KamLAND-Zen:2016pfg})} \\
        &1.07 \times 10^{26} \mbox{yr~~(\ce{^{136}Xe},~\cite{GERDA:2020xhi})}
     \end{aligned} 
     \right. .
   \end{equation}
   By inserting the appropriate values for the NMEs, phase space factors and for the other quantities, it is possible to obtain a lower bound on the compositeness scale $\Lambda$ as a function of 
   the heavy composite Majorana neutrino mass $M_N$ from the inequality
   \begin{eqnarray}
   \label{NEWbound}
      \Lambda \geq \frac{g_*}{2^{3/4}} \sqrt{\frac{\eta_L \, g_A}{ G_F \cos \theta_c}} \left( \frac{m_p}{M_N} \right)^{\frac{1}{4}}
      \left( G_{01} \, |\mathcal{M}^{0N}|^2 \,  T_{1/2}^{\textrm{exp.}} \right)^{\frac{1}{8}} \, ,
      \nonumber \\
   \end{eqnarray}
   upon requiring  $T_{1/2}^{\textrm{CI}} \geq T_{1/2}^{\textrm{exp.}}$.
   We set $\eta_L$ to unity, as commonly performed in the phenomenological and experimental collider-based analyses.
   The resulting bound for the \ce{^{136}Xe} case is shown in Fig.~\ref{FigBound} in the model parameter-plane $(M_N, \Lambda)$.
   The (red semi-transparent) band is obtained by varying the NME in the range 
   $(72.6, 186)$, which correspond to minimum (IBM model, \cite{Barea:2015kwa}) and maximum (QRPA model, \cite{Hyvarinen:2015bda}) values for $\mathcal{M}^{0N}$; other calculations lead to intermediate values~\cite{Engel:2016xgb,Menendez:2017fdf} (NSM model).
   The uncertainty on the phase space factor $G_{01}$ is practically negligible~\cite{Kotila:2012zza}.  The half-life limit of \ce{^{76}Ge} is tighter, but the corresponding \bb\ bound in the $(M_N, \Lambda)$ plane is less constraining, mainly due to a smaller value of the phase space factor.


In the left panel of Fig.~\ref{FigBound} we compare the \bb\ bound with the exclusion limits provided by the LHC analysis (Run 2) searching for the composite neutrino within the same
   Lagrangian model \cite{Sirunyan:2017xnz}; the excluded regions have to be understood below the curves.
   One can see that the \bb\ is rejecting portions of the parameter space $( M_N,\Lambda)$ still allowed by the CMS data (blue dots). 
   In particular, for $\Lambda=M_N$ the LHC search~\cite{Sirunyan:2017xnz} excludes masses $M_N<4.6$\,TeV, while the \bb\ search masses $M_N<(7.3-8.8)$ TeV depending on the selected value for the NME.
   It is worth noticing that the \bb~bound performs better also in the low-mass region, where the Run-2 analysis loses sensitivity due to less energetic particles in the final state. 

   In a similar way, it is possible to foresee the sensitivity on the compositeness scale coming from the future searches for \bb\ and the projection study for the High-Luminosity LHC (HL-LHC),
   that will operate with a centre-of-mass energy of 14 TeV and an integrated luminosity of 3 ab$^{-1}$.
   The next generation of \bb\ experiments aims at sensitivities for the half-life of more than $10^{27}$\,yr, up to $10^{28}$\,yr.
   We adopt
   \begin{equation}
     \taubb > 10^{28} \mbox{yr~~(\ce{^{136}Xe} projection,~\cite{nEXO:2017nam})} \, ,
   \end{equation}
   when extracting the \bb\ bound with Eq.~\eqref{NEWbound}.
   On the collider side, we use the projection study for the search of the composite neutrino included in the recent Yellow Report CERN publication~\cite{CMS-PAS-FTR-18-006,CidVidal:2018eel}. 
   
   The future exclusions limits  are given in the right panel of Fig.~\ref{FigBound}. 
   Here, the situation is rather different with respect to the current exclusion limits: the HL-LHC allows to discard a larger portion of the parameter space than the \bb\, both in the low-mass region
   and up to $M_N \simeq 6.5$ TeV.
   Nevertheless, the bound from the $^{136}$Xe decay still gives the strongest exclusion limit in the high-mass region, improving the collider-driven bound on the composite-neutrino mass at
   $\Lambda=M_N$ from $M_N=8.0$ TeV to $M_N=(11.5-12.6)$ TeV. 
   
   Let us summarize our findings.
   Excited fermions are actively searched for at collider facilities, and the most recent bounds on their masses and the compositness scale have been pushed to the multi-TeV range
   by the ATLAS and CMS Collaborations. In this work, 
   we complemented the collider-driven exclusion limits on the composite-excited neutrino by considering the low-energy nuclear decay $\ce{^{136}Xe \to ^{136}Ba + 2e^-}$. Here, upon assuming that the \bb\ process is mediated by the very same composite neutral lepton, which interacts via contact interactions with SM quarks and leptons, we can extract stringent bounds on the model parameter space $(M_N,\Lambda)$, as shown in Fig.~\ref{FigBound}.
   We find that \bb-driven bound is highly competitive with the current CMS search, and it remains fairly competitive even with the HL setting of the LHC, especially in the high mass region.
   Despite the \bb\ bound genuinely applies to composite Majorana neutrino only, the indication on the excited fermion mass and compositeness scale can be anyhow taken as orientation for other
   excited states, namely quarks and charged leptons.

\section*{Acknowledgments}

    This work is partially supported by Research Grant Number 2017W4HA7S ``NAT-NET: Neutrino and Astroparticle Theory Network'' under the program PRIN 2017 funded by the Italian ``Ministero dell’Istruzione, dell'Universit\`a e della Ricerca'' (MIUR).

   \bibliography{new_ref}

\end{document}